\documentclass[aps,
preprint,
superscriptaddress,groupedaddress]{revtex4} 
\usepackage{latexsym}       
\usepackage{natbib}
\usepackage{graphicx}  
\usepackage{anysize} 
\usepackage{array}
\usepackage{float}
\usepackage[T1]{fontenc}           
\begin{document}

\title{Epidemic spreading in multiplex networks influenced by opinion
  exchanges on vaccination}

\author{L. G. Alvarez-Zuzek} \email{lgalvere@mdp.edu.ar}
\affiliation{Instituto de Investigaciones F\'{i}sicas de Mar del Plata
  (IFIMAR-CONICET), Facultad de Ciencias Exactas y Naturales,
  Universidad Nacional de Mar del Plata, D\'{e}an Funes 3350, Mar del
  Plata, Argentina}

\author{Cristian E. La Rocca} \affiliation{Instituto de
  Investigaciones F\'{i}sicas de Mar del Plata (IFIMAR-CONICET),
  Facultad de Ciencias Exactas y Naturales, Universidad Nacional de
  Mar del Plata, D\'{e}an Funes 3350, Mar del Plata, Argentina}

\author{J. R. Iglesias} \affiliation{Programa de Pós-Graduação em
  Economia, Escola de Gestão e Negócios, UNISINOS, 93022-000, São
  Leopoldo, RS, Brazil} \affiliation{Instituto Nacional de Ci\^encia e
  Tecnologia de Sistemas Complexos, CBPF, Rio de Janeiro, RJ, Brazil}
 \affiliation{Instituto de Investigaciones F\'{i}sicas de Mar del Plata
  (IFIMAR-CONICET), Facultad de Ciencias Exactas y Naturales,
  Universidad Nacional de Mar del Plata, D\'{e}an Funes 3350, Mar del
  Plata, Argentina}

\author{L. A. Braunstein} \affiliation{Instituto de Investigaciones
  F\'{i}sicas de Mar del Plata (IFIMAR-CONICET), Facultad de Ciencias
  Exactas y Naturales, Universidad Nacional de Mar del Plata, D\'{e}an
  Funes 3350, Mar del Plata, Argentina}

\begin{abstract}

Through years, the use of vaccines has always been a controversial
issue. People in a society may have different opinions about how
beneficial the vaccines are and as a consequence some of those
individuals decide to vaccinate or not themselves and their
relatives. This attitude in face of vaccines has clear consequences in
the spread of diseases and their transformation in
epidemics. Motivated by this scenario, we study, in a simultaneous
way, the changes of opinions about vaccination together with the
evolution of a disease. In our model we consider a multiplex network
consisting of two layers. One of the layers corresponds to a social
network where people share their opinions and influence others
opinions. The social model that rules the dynamic is the M-model,
which takes into account two different processes that occurs in a
society: persuasion and compromise. This two processes are related
through a parameter $r$, $r<1$ describes a moderate and committed
society, for $r>1$ the society tends to have extremist opinions, while
$r=1$ represents a neutral society. This social network may be of real
or virtual contacts. On the other hand, the second layer corresponds
to a network of physical contacts where the disease spreading is
described by the SIR-Model. In this model the individuals may be in
one of the following four states: Susceptible ($S$), Infected($I$),
Recovered ($R$) or Vaccinated ($V$). A Susceptible individual can: i)
get vaccinated, if his opinion in the other layer is totally in favor
of the vaccine, ii) get infected, with probability $\beta$ if he is in
contact with an infected neighbor. Those $I$ individuals recover after
a certain period $t_r=6$. Vaccinated individuals have an extremist
positive opinion that does not change. We consider that the vaccine
has a certain effectiveness $\omega$ and as a consequence vaccinated
nodes can be infected with probability $\beta (1 - \omega)$ if they
are in contact with an infected neighbor. In this case, if the
infection process is successful, the new infected individual changes
his opinion from extremist positive to totally against the vaccine. We
find that depending on the trend in the opinion of the society, which
depends on $r$, different behaviors in the spread of the epidemic
occurs. An epidemic threshold was found, in which below $\beta^*$ and
above $\omega^*$ the diseases never becomes an epidemic, and it varies
with the opinion parameter $r$.

\end{abstract}

\maketitle

\section*{Introduction}
\label{intro}

In 1796 Edward Jenner invented and tested a vaccine against the
smallpox, an illness that had a very high index of mortality in the
18TH century~\cite{cdc_small}. The idea of Jenner was so successful
that nowadays smallpox is practically eradicated and after this
pioneering essay different vaccines were elaborated to prevent a long
list of infectious diseases, from poliomyelitis to influenza. However,
vaccines may present some lacks of efficiency and also some collateral
effects. For example, in recent years some publications wrongly
associated vaccination with autism~\cite{cdc,Bel03}. In spite of
the overwhelming scientific evidence that such correlation is not
actual, the belief that the results of vaccination could be worse than
the illness itself spread through social networks and generated groups
and movements against vaccination. Sometimes these groups are also
related to some religion beliefs and/or rightists political tendencies
or candidates. The debate about the efficiency of vaccination and its
possible risks is then a very actual debate and a typical example of
propagation of opinions, for and against vaccination. Thus,
considering that opinions and contagion spreads in different ways, we
will perform this study on a Network on Network. In recent years the
study of complex Network of Networks (NoN) has been a subject of great
interest for the scientific community, due to the large number of real
word systems that can be mimic and study using these kind of
topological structures \cite{Jianxi_10,Perc_15,Bocca_14,Kiv_13}. A NoN
is a system formed by single networks interacting through external
connections between them. Many researches on NoN were focused in the
study of cascade of failures \cite{Bul_01,Bru_01,dim_16}, propagation
of epidemics \cite{Arenas_16,Cozzo_13,Saumell_12,Pires_17}, and
opinion dynamics
\cite{Fort_01,Halu_13,Diakonova_14,Diakonova_16,Gal_08} due to the
ubiquitous of these processes that are present in the real
scenarios. In particular, we are interested in processes that develop
on NoN in which nodes belonging to different networks represents the
same entities. This type of NoN are usually called {\it multiplex}
networks. Epidemic spreading models have been particularly successful
in understanding and predicting an epidemic outbreak and its period of
extinction. Also, some models have incorporated a factor of human
behavior, by considering the information and sources of information
that individuals must handle, rational decisions and behavioral
changes, in order to reach a more comprehensive understanding about
the epidemic spreading \cite{Funk_10}. A commonly-used model for
reproducing spreading diseases dynamics in networks is the
susceptible-infected-recovered (SIR) model
\cite{Ander_91,Bailey_75,New_05,Pastor_15,Mil_02}. This model has been
successfully used to reproduce non recurrent diseases such as the H5N5
flu or the Severe Acute Respiratory Syndrome (SARS)
\cite{Col_07}. Besides, it has been extensively studied under the
topology of multiplex networks
\cite{Dickison_12,Mar_11,Yag_13,Buono_14}. The model groups the
population of individuals to be studied into three compartments
according to their state: the susceptible ($S$), the infected ($I$),
and the recovered ($R$). When a susceptible node is in contact with an
infected node it becomes infected with an intrinsic probability
$\beta$, which we called the virulence of the disease, and after a
period of time $t_r$ it recovers and becomes immune. Usually, the type
of disease that this model describes has a period of infection that
lasts for six or seven days on average, flu, for example.

The study of these models in real and synthetic networks
\cite{tizzoni2012real,Valdez_Ebola} have allowed researchers to
develop different mitigation strategies for decreasing the impact of
diseases on healthy populations
\cite{wang_16,Val_13,Alvarez-zuzek_15,Buono_15}. These studies have
been used in government policies to design vaccination campaigns. For
instance, for seasonal diseases, such as influenza, vaccination
campaigns are scheduled to begin before the epidemic spreads and in
general this strategy is very effective \cite{Buono_15}. Another
strategy of prevention is the isolation for a certain period of time
of individuals with infectious symptoms to prevent the spreading
\cite{Val_13,Alvarez-zuzek_15}. Note that these scenarios are
particularly interesting for epidemics spreading and the question that
motivates this work is how the spread of the disease is influenced and
co-evolves with the social context.  Within the context of social
phenomena, many empirical investigations show the importance of social
influence in the formation of people's opinions. It is argued that two
interacting partners may exert social pressure to change their
attitudes approaching their opinions \cite{Fest_50}. This particular
social mechanism is named {\it compromise}
\cite{Fort_01,Weis_02,BenNaim_D03,BenNaim_A03}. A less explored
mechanism of social interactions is the {\it persuasion}
\cite{Mye_82,Isen_86,Lau_98,Mas_13b}. Myers \cite{Mye_82} observed in
group discussion experiments that when two individuals talk, they do
not only state their opinions, but they also discuss about the
arguments that support their opinions. If they hold the same opinion,
they could strength it by persuading each other with new arguments or
reasons, becoming more extreme in their believes. In this context, La
Rocca {\it et. al} \cite{Larocca14} proposed and studied a model that
explains the phenomena of polarization in a population of individuals
that evolve under pairwise interactions, by implementing those two
main social mechanisms of opinion's formation, i.e., compromise and
persuasion \cite{Mas_13,Mas_13b,bal_15}. This model, denoted as the
M-model, has $2M$ different states describing the spectrum of possible
opinion orientations on a given issue, from totally against (state $x
= -M$) to totally in favor ($x = M$), with some moderate opinions
between these extreme values.

The study of opinion dynamics on NoN is relatively new
\cite{Bocca_14}. Alvarez-Zuzek {\it et al.} \cite{Alvarez_2016}
investigated the interaction between two social dynamics, one for
opinion formation and the other for decision making, on two
interconnected networks. The dynamics for opinion formation
corresponds to the M-model proposed in \cite{Larocca14}, and the
decision making dynamics is akin to the Abrams-Strogatz (AS) model
\cite{AS_03,vazquez_10} originally introduced to study language
competition, where agents can choose between only two possible options
($x = \pm 1$). In this model each agent may change its decision by a
mechanism of social pressure, in which the probability of switching
his present choice increases non-linearly with the number of neighbors
that have the opposite opinion. They concluded that under certain
parameters of the system, one model prevails over the other and
dominates the behavior of the system.

The goal of the present contribution is to investigate the effect of
the dynamic of opinion formation on vaccination on the evolution of a
given disease, for instance the flu. Thus, we will study the
propagation of a disease in a population where all the individuals are
continuously debating about getting vaccinated, considering that a
susceptible individual is vaccinated if he is completely convinced
about the benefits of the vaccine. However if after being vaccinated
he catch the disease he becomes completely against the vaccination.
For this purpose, and because the two processes occur on the same
group of individuals, we studied the SIR model with vaccination and
the M-model in a multiplex system composed by two networks. Both
dynamics take place in different layers and co-evolve. Susceptible
individuals become vaccinated if they acquire the state $M$ in the
other network, while the vaccinated individuals acquire the state $-M$
if they get infected.

The paper is organized as follows: in the next section we expose the
model presented in its extended form. In Section 3, we present the
simulation results and Section 4 is devoted to discussion and
conclusions.

\section*{The Model}

We are interested in studying how the propagation of diseases is
influenced by the opinion formation of individuals in favor or against
of getting vaccinated. The opinions will be formed and/or modified
through the interaction and exchange of ideas with other individuals,
which have their own opinion and co-evolves with the health condition
of those individuals. In this way, the group of individuals develop a
dynamic of formation of opinions in which individuals interact
expressing opinions about the importance or not of being
vaccinated. When an individual has a fully positive opinion about the
vaccine, he acts accordingly and gets vaccinated. In our model we do
not consider parental decisions on children, so the opinion on
vaccination motivates just individuals, not family groups. While the
process of spreading a disease generally requires face-to-face
physical contact, the process of formation of opinions is more
flexible because opinions can be transmitted via other media: phone
calls, online social networks, video conference and instant messaging
services, etc.

To model the spread of the disease in layer $A$ we use a variation of
the SIR model in which a new stage of healthy vaccinated individual
($V$) is incorporated. Let's recall that vaccinated individuals share
opinion $M=+2$. i) An individual $S$ in contact with an infected
individual $I$ becomes infected with a probability $\beta$ (the
infectivity of the disease). ii) However, as the vaccine does not
guarantee $100\%$ protection a vaccinated individual ($V$) can become
infected with a probability $(1-\omega) \beta$, where $\omega$ is the
efficiency of the vaccine. iii) An infected individual recovers after
a period of time $t_r$, and we assume he acquires immunity. If the
vaccinated agent gets infected, he changes to the opposite opinion
becoming an extremist against the vaccine ($M=-2$). This may be an
extreme behavior and probably people are not so extremists, but our
objective here is to describe the frustration of a vaccinated agent
after acquiring the infection. Notice that, if we let this frustrated
agent go to an intermediate opinion, the influence on the epidemics
evolution is almost irrelevant.

For the process of opinion formation in network $B$ we use the $M$
model \cite{Larocca14}. This model explains the phenomena of
polarization in a population of interacting individuals and two main
processes are involved: compromise and persuasion. We consider $M=2$,
being $M$ (totally in favor) and $-M$ (totally against) the extremist
cases and the intermediate cases correspond to the states of moderate
opinion. In our model we considered that only one individual, the
$i-$agent, can change his opinion, assuming that the other one, the
$j-$agent has enough arguments to convince or change the opinion state
of the first individual. Then, the rules of the opinion model are:

A node $i$ is chosen and it can change its opinion state after
interaction with a neighbor $j$. If their respectively opinion states
are $x_i$ and $x_j$, we proceed as follows:

\begin{itemize}

\item If both individuals have the same opinion orientation
  (i.e. $x_ix_j>0$), then with probability ${\bf p}$: $x_i= \pm 1
  \rightarrow x_i= \pm 2$.

In case that $i$ is already in an extremist state ($x_i\pm 2$) it
remains extremist.

\item If both individuals have different opinion orientations
  (i.e. $x_ix_j<0$), then with probability ${\bf q}$\\
  
  $x_i=\pm 1 \rightarrow x_i=\mp 1$,\\

  $x_i=\pm 2 \rightarrow x_i=\pm 1$.

However, one assumes that if node $i$ is a vaccinated agent, he keeps
his opinion (and his vaccinated state) even when interacting with
neighbor $j$ having an opposite sign opinion.

\end{itemize}

If two nodes have the same opinion orientation, one of them becomes
more extremist with probability $p$, but if they have different
opinion orientations one of the individuals becomes more moderate with
probability $q$. For simplicity, we consider $p+q=1$ and define the
ratio $r=p/q$. In our model an individual $S$ becomes $V$ if in layer
$B$ he reaches opinion $2$. On the other hand, if an individual $V$
becomes $I$ with probability $\beta (1-\omega)$, then in layer $B$ he
changes his opinion to $-2$. Notice that even when the recovered
individuals becomes inactive in the layer $A$, they are still active
in the layer $B$.

\begin{figure}[H]
  \vspace{1cm}
  \begin{center}
    \includegraphics[scale=0.55]{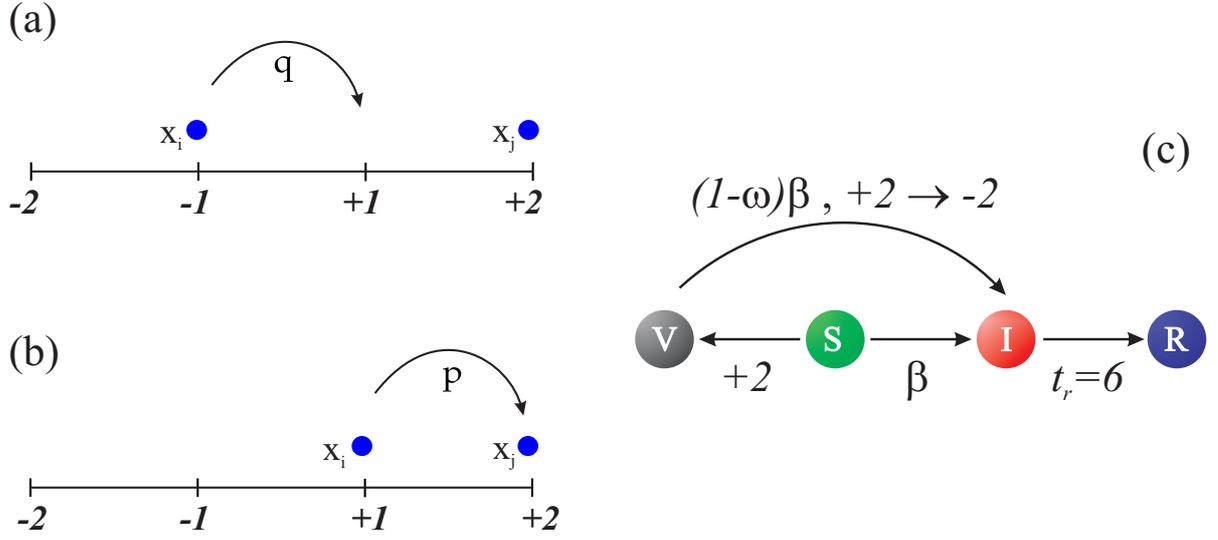}
  \end{center}
  \caption{Simplified scheme of opinions and epidemics dynamics: Left
    figures illustrate the opinion dynamics, when two nodes have
    opinion states with different sign, one of them approaches its
    state to the opinion of the other with a probability $q$ (a),
    whereas if the sign is the same the node reinforces its opinion
    with probability $p$ (b). Right figure illustrate the contagion
    dynamics, a susceptible individual $S$ (green) is infected (red)
    with probability $\beta$ and after a time $t_r$ he recovers
    (blue). A $S$ becomes vaccinated $V$ (gray) when he acquires a
    state of opinion $2$, but then he can become $I$ with a
    probability $ (1- \omega) \beta $, changing his opinion to $-2$.}
  \label{fig_esq.eps}
\end{figure}

In Fig. \ref{fig_esq.eps} we show a scheme of the rules of the
dynamics of the whole system. An individual with opinion state $x_i$
whose neighbor with state $x_j$ has an opinion with different
orientation approaches to the opinion of the neighbor with probability
$q$ (Fig. \ref{fig_esq.eps}a)), whereas if both individuals have the
same orientation of opinion i-agent reinforces his opinion with
probability $p$ (Fig. \ref{fig_esq.eps}b)). Concerning the
contagion (Fig. \ref{fig_esq.eps}c)), an individual $S$ (green)
becomes $I$ (red) with a probability $\beta$ and after a time $t_r$
goes to the recovered state $R$ (blue). An agent $S$ is vaccinated $V$
(gray) when acquiring the opinion state $2$, and in contact with an
infected individual can get infected with a probability $(1- \omega)
\beta $, where $\omega$ ($ 0 \leq \omega \leq 1$) is the efficiency
of the vaccine. If the $V$ agent is infected he loses its confidence
in the vaccine and thereby changes his state of opinion from $2$ to
$-2$.

\section*{Simulation Results}

We study the model described in the previous section by means of
extensive Monte Carlo simulations with synchronous update using a
two-layer network of the same size $N=10^5$. Nodes in each layer
represent the same agent, thus we connect through an external link a
pair of nodes, each from different layer, allowing only one interlink
by node. We construct each layer using the Molloy-Reed algorithm
\cite{Mol_01} considering the Erd\H{o}s-R\'enyi (ER) \cite{Erd_01}
degree distribution with $\langle k \rangle = 4$. The propagation of
the disease takes place in layer $A$ and we fix the recovery time in
$t_r=6$, which is in days the characteristic period of infection for a
flu. Layer $B$ is the social network, where the M-model rules the
dynamic, with $M=2$. As initial conditions we use for the layer $B$ an
uniform distribution for the densities of opinion, i.e., the same
initial probability $P_{+2,+1,-1,-2} = 1/4$. In layer $A$ we have
initially only one agent infected, which is considered the patient
zero and whose opinion is chosen at random between the four possible
opinion states, a fraction $1/4$ of the agents are vaccinated ones, as
a consequence of their opinion state $+2$, and the rest are
susceptible. We chose one source node of infection because this is
the standard approach used by epidemiologists where most outbreaks
starts with one person. At each time steps, we first let evolve the
epidemic dynamic and then the opinion process. In layer A, we allow
all the infected individuals to infect each one of their susceptible
neighbors with probability $\beta$ and the vaccinated neighbors with a
probability $(1-\omega)\beta$. Then, in the opinion layer, we iterate
over all the individuals and give each one of them the chance to
interact with only one of its neighbors. This neighbor is chosen among
those who can change the individual opinion. In case there is no
neighbor that can change the opinion, nothing happen. Finally, we
update all the opinions and epidemic states at the next time
step. Notice that those infected individuals who had $t_r$ time step
to spread the disease recover and those susceptible individuals whose
opinion change to $+2$ change into the vaccinated state. All numerical
results correspond to an average over $10^5$ independent realizations.

We concentrate in the steady state of the system which is reached when
the number of infected nodes becomes zero, regardless of whether
consensus was reached in the opinion network. Then, the magnitudes to
be studied are the fraction of recovery nodes ($R$), the fraction of
vaccinated nodes ($V$), the duration time of the epidemic ($\tau$) and
the magnetization of the opinions ($Mag$). Notice that at any time
$S+I+R+V=1$ and $Mag = \sigma_+ - \sigma_-$, where $\sigma_{\pm}$ is
the fraction of nodes with positive (negative) opinion state. We only
take into account those realizations in which an epidemic occurs,
i.e., the total number of recovered individuals in the final state is
greater than a cutoff $200$ for a system size of $N=10^5$
\cite{Lag_02}.

\begin{figure}[H]
  \begin{center}
    \includegraphics[scale=0.55]{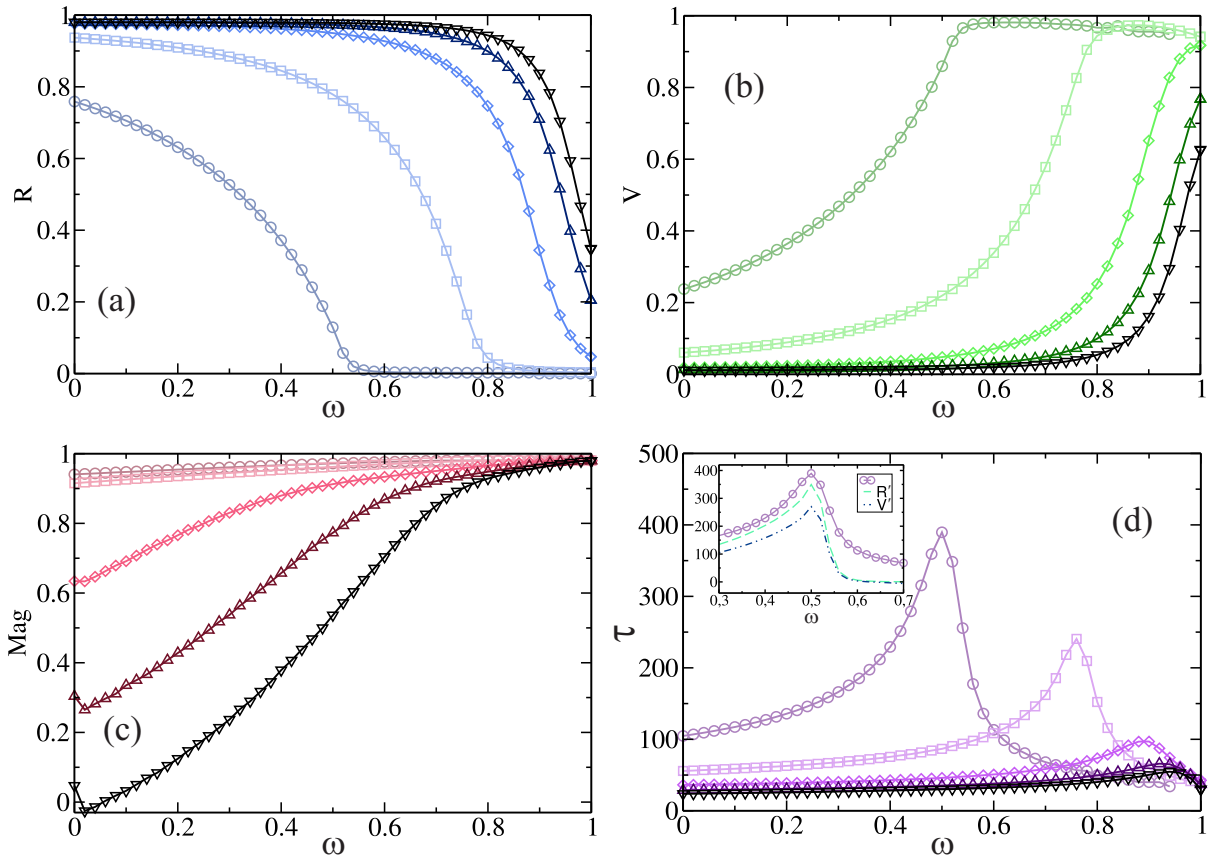}
  \end{center}
  \caption{(a) Fraction of recovery individuals $R$, (b) Fraction of
    vaccinated individuals $V$, (c) Magnetization of the opinions
    $Mag$ and (d) the duration time of the epidemic $\tau$, as a
    function of the efficiency of the vaccine $\omega$. Inset: $\tau$
    (solid line), the derivative of $R$ (dashed line) and the
    derivative of $V$ (dot dashed line) as a function of the
    efficiency $\omega$ and $\beta = 0.1$. From the inset it is clear
    that the maximum duration of the epidemics corresponds to
    inflection points in the number of recovered and vaccinated
    agents. In all cases we set $t_r=6$ and $r=0.1$ for $\beta=0.1$
    ($\bigcirc$), $0.2$ ($\Box$), $0.4$ ($\diamond$), $0.6$
    ($\bigtriangleup$) and $0.8$ ($\bigtriangledown$). All numerical
    results correspond to an average over $10^5$ independent
    realizations.}
    \label{Fig1}
\end{figure}

Now, we will present in further details $R$, $V$, $\tau$ and $Mag$ as
a function of the efficiency of the vaccine $\omega$, for different
values of the parameters $r$ and $\beta$. In Fig. \ref{Fig1} we
present the case $r=0.1$, that mimics a scenario in which the
population's opinion settles in a centralized state where the
compromise process dominates.

In Fig. \ref{Fig1}(a) we show the total fraction of recovery nodes as
a function of $\omega$ for different values of $\beta$. We can observe
that for certain values of $\beta$, as $\omega$ increases the fraction
$R$ decreases and above a value $\omega^*$, which is a threshold for
the efficiency of the vaccine, the system does not present an
epidemic phase and corresponds to the inflection point of the
curve. This is because as the vaccine becomes more effective, more
people remain vaccinated and the propagation of the disease slows
down. An efficiency above the threshold $\omega > \omega^*$ is
enough to ensure that an epidemic will not develop, such as the case
of low values of $\beta$. For example for $\beta=0.1$ we need an
efficiency of at least $55\%$ in order to avoid the epidemic. On
the other hand, above a certain value of $\beta^*$ the propagation of
the disease is enhanced and it is impossible to prevent an
epidemic. Even for $\omega=1$ there will be a macroscopic number of
recovery individuals in the steady state. This is the case for the
values $\beta=0.4$, $0.6$ and $0.8$.  Fig.~\ref{Fig1}(b) and (c) shows
the fraction of vaccinated nodes and the magnetization of the opinions
as a function of $\omega$. For all the values of $\beta$ we can see
that both magnitudes increases with $\omega$. This is consistent with
the fact that as the vaccine becomes more efficient, more people will
agree to get vaccinated.

For $r=0.1$ the compromise is higher than the persuasion and as a
consequence agents tend to have moderate opinions (in favor or
against). However, when an agent is vaccinated, his opinion ($2$)
remains fixed producing an attractive effect towards positive opinion
and he will only change his opinion if he gets infected. As can see
from Fig. \ref{Fig1}(c), for low values of efficiency $Mag$ decreases
as $\beta$ increases, for example, for $\beta=0.8$ agents opinion are
in a polarized state ($Mag=0$). This behavior is due to the fact that
as $\beta$ increases more vaccinated agents gets infected, so their
opinion change from $2$ to $-2$, which means that more people become
extremist against the vaccine. On the other hand, for an efficiency
close to one, the opinion of the system is in average almost
completely in favor of the vaccine, reaching a consensus where all the
agents have the same opinion. Because the efficiency of the vaccine is
high the vaccinated agents stay pinned in the opinion $+2$, pushing
all the agents to adopt their opinion. Notice that in this scenario
the convergence time of both dynamics are similar, $i.e$, the time
that it takes to the disease to propagates all over the population
allows the people to reach consensus in favor of the vaccine.

In Fig. \ref{Fig1}(d) we can see the duration time of the epidemic as
a function of $\omega$. As we can observe as $\omega$ increases more
nodes are vaccinated and as a consequence the duration of the disease
increases. Around the threshold $\omega^*$ the time of the epidemic
exhibits a peak and then decreases rapidly. This is consistent with
the fact that as $\omega$ increases the number of $R$ decreases, which
means that it is hard to spread the disease and therefore $\tau$
increases. The time of duration of the epidemic reaches a maximum at
$\omega^*$ and then decreases because the spreading of the disease is
diminish (there is no epidemic). Note that for $\omega >\omega^*$ the
majority of the agents in the system are vaccinated. We added an inset
in Fig. \ref{Fig1}, (as well as in the following ones), comparing the
duration of the epidemics with the derivatives of the number of
recovered ($R$) and vaccinated ($V$) agents as a function of the
efficiency. It is possible to see that there is an inflection point (a
maximum in the derivatives) when the duration of the epidemics is
maximum, meaning that the number of recovered and vaccinated agents
increase at a lower rate when the efficacy of the vaccine is high than
when the efficacy is low, going trough a maximum rate when the
duration of the epidemics is maximum.

\begin{figure}[H]
  \begin{center}
    \includegraphics[scale=0.55]{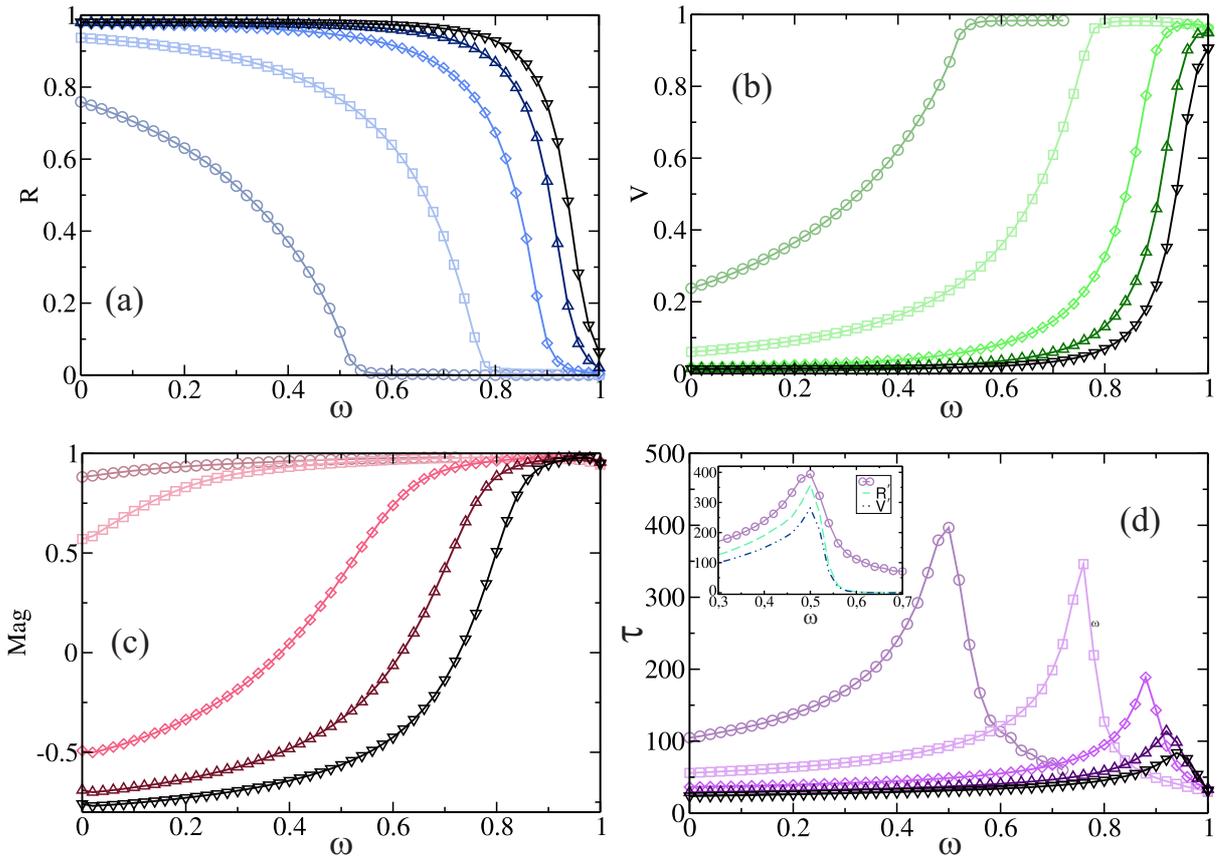}
  \end{center}
  \caption{ (a) Fraction of recovery individuals $R$, (b) fraction of
    vaccinated individuals $V$, (c) Magnetization $Mag$ and (d) the
    duration time of the epidemic $\tau$, as a function of the
    efficiency of the vaccine $\omega$. Inset: $\tau$ (solid line),
    the derivative of $R$ (dashed line) and the derivative of $V$ (dot
    dashed line) as a function of the efficiency $\omega$ and $\beta =
    0.1$. In all cases we set $t_r=6$ and $r=1$ for the same values of
    $\beta$ and symbols used in Fig. \ref{Fig1}. All numerical results
    correspond to an average over $10^5$ independent realizations.}
  \label{Fig2}
\end{figure}

In Fig. \ref{Fig2} we show the case of $r=1$, which mimics a neutral
society where the probability of compromise and persuasion are equal
($p=q=0.5$). Fig.~\ref{Fig2}(a) shows $R$ as a function $\omega$. We
can observe that an efficiency threshold exists (for low values of
$\beta$) below which the diseases never becomes an epidemic. While for
higher values of $\beta$ there is always an epidemic phase. In
Fig.~\ref{Fig2}(c) we show $Mag$ as a function of $\omega$. From the
plot we can see that for low values of $\omega$ the magnetization
decreases as $\beta$ increases, while for an efficiency close to one
the system reaches consensus in favor of the vaccine. Besides, for low
values of $\omega$ and high values of $\beta$, $Mag$ becomes negative
because of the persuasive effect, which is not negligible. Agents with
negative opinion are less likely to change their opinion. Also, since
the vaccine is not much effective, vaccinated agents gets infected,
thus their opinions change to negative and this contributes to a
negative magnetization. On the other hand, we can see that for large
values of $\omega$ the convergence time of both dynamics are similar,
the $Mag$ in the opinion dynamic is close to one, thus is close to the
consensus state. The opposite occurs for the other values of $\omega$,
where the model of opinions is far from consensus.

\begin{figure}[H]
  \begin{center}
    \includegraphics[scale=0.55]{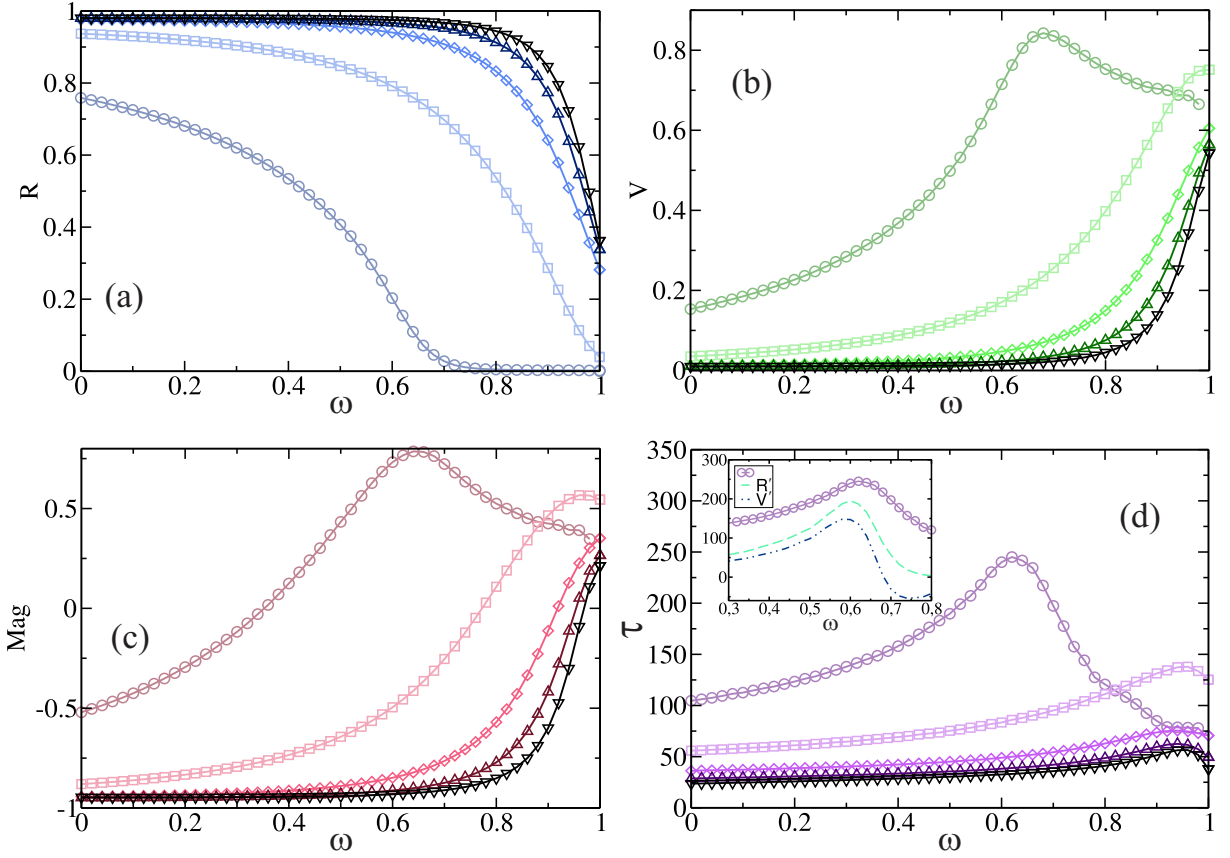}
  \end{center}
  \caption{(a) Fraction of recovery individuals $R$, (b) fraction of
    vaccinated individuals $V$, (c) Magnetization $Mag$ and (d) the
    duration time of the epidemic $\tau$, as a function of the
    efficiency of the vaccine $\omega$. Inset: $\tau$ (solid line),
    the derivative of $R$ (dashed line) and the derivative of $V$ (dot
    dashed line) as a function of the efficiency $\omega$ and $\beta =
    0.1$. In all cases we set $t_r=6$ and $r=10$ for the same values
    of $\beta$ and symbols used in Fig. \ref{Fig1}. All numerical
    results correspond to an average over $10^5$ independent
    realizations.}
  \label{Fig3}
\end{figure}

Now we will show the case of $r=10$, which represents an extremist
society where persuasion dominates the process of opinion
formation. In this case, agents with extremist opinion tend to
convince agents with moderate opinion to become extremists. From
Figs. \ref{Fig3}(a) and (d) we can see that the behavior of $R$ and
the time duration of the epidemic $\tau$ are qualitative the same that
for the two previous cases studied of $r$ with different
$\beta^*$. Agents become extremist in their opinions and those who are
against the vaccine have a small probability to be vaccinated. Then
the disease spreads is promoted among the non vaccinated agents, which
are an important fraction of the population. In Figs. \ref{Fig3}(b)
and (c) we can observe that $V$ and $Mag$ increase as $\omega$
increases, and for the cases below $\beta^*$ (as $\beta=0.1$) both
magnitudes increase with $\omega$ until reaching a peak around
$\omega^*$ after which these magnitudes decreases. This is due to the
fact that above the point $\omega^*$, the time of the epidemic
decreases as $\omega$ increases, and there is not enough time to
convince the negative opinion agents to vaccinate.

As the persuasion is higher than compromise ($p=0.91$ and $q=0.09$),
agents tend to remain with extremist opinions, against or in favor of
the vaccine. The attractor effect that generate the vaccinated agents
in the opinion state is now hidden by the persuasive effect. The
persuasive effect moves agents to the extreme opinions. When a
vaccinated agent is infected, its opinion becomes a negative extremist
and in this extremist society he will rarely change his opinion. For
this reason, for all values of $\beta$ and low values of $\omega$ the
$Mag$ is always negative. On the other hand, for high values of the
efficiency the average opinion of the system can be in favor or
against depending of the virulence, but in general it is polarized and
as a consequence, $Mag$ closed to zero. Even for the case in which
$\omega=1$ there still some agents that are against the vaccine. This
is due to the fact that almost all agents that began with negative
opinion remains in that state. Notice that the epidemic dynamics is
faster than the dynamics of opinions -the convergence time is higher
in layer $B$- making that the population never reach a consensus of
opinions. \\

\begin{figure}[H]
  \begin{center}
    \includegraphics[scale=0.5]{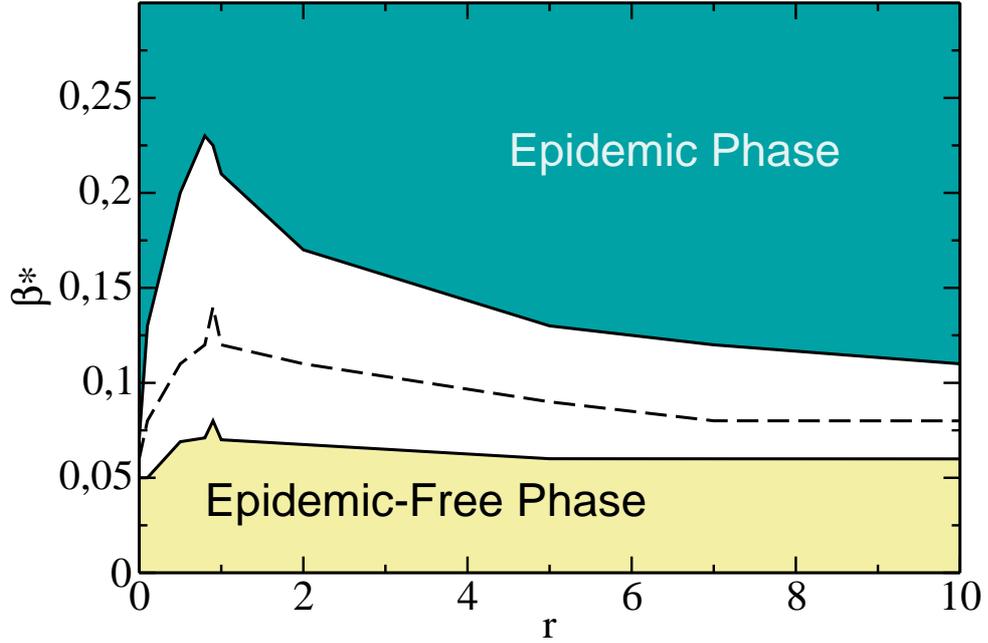}
  \end{center}
  \caption{Threshold epidemic value $\beta^*$ as a function of $r$,
    with $t_r = 6$, $\omega = 1$ and as initial conditions $1\%$,
    $10\%$ and $25\%$ of vaccinated nodes, from bottom to top. The
    maximum value is closed to $r \approx 1$, which is a neutral
    society.}
  \label{Fig4}
\end{figure}

We mentioned before that there is a threshold $\beta^*$ above which
the system always stay in an epidemic regime, independently of the
efficiency of the vaccine, $\omega$. In Fig. \ref{Fig4} we show
$\beta^*$ as a function of $r$ for $t_r=6$ and $\omega=1$. We set
$\omega = 1$, so that the vaccine is $100\%$ effective, in order to
know how strong the virulence of the disease has to be to win the best
vaccination scenario. In addition, we also study this scenario for
different values of initial vaccinated nodes -$1\%$, $5\%$ and $25\%$-
to see how the initial conditions impacts on the evolution of both
dynamics. As we can see from Fig.~\ref{Fig4}, for a certain value of
$r$, $\beta^*$ decreases as the initial vaccinated nodes
decreases. This is consistent with the fact that having fewer initial
vaccinated nodes causes the disease to spread more easily, so that
less virulent diseases could become epidemic. As we can observe that
the maximum values are around $r \approx 1$, which means that a
neutral society is optimal to prevent an epidemic. In a society with
$r<<1$, compromise dominates the process of opinion formation and the
agents tend to have a moderate opinion. This prevents those moderate
agents from becoming extremists in favor of vaccination. The disease
spreads through the non vaccinated agents very easily, even when the
virulence is small. On the other hand, in a society with $r>>1$,
persuasion dominates the process of opinion formation and the agents
tend to adopt extremists opinion. All extremists agents in favor of
vaccination will be vaccinated, but those agents with a negative
opinion (against the vaccine) have a small probability to be
vaccinated because they will hardly change their opinion. In this
case, the disease spreads very easily over these agents, which are an
important fraction of the population. In a neutral society it is more
likely for an agent with a positive moderate opinion to become an
extremist in favor of the vaccine than in the case $r<<1$, and it is
more likely for an agent against the vaccine to change his opinion in
favor than in the case $r>>1$. For this reason, $\beta^*$ is higher
than in the other cases, because it is easier to convince people to
get vaccinated, making more difficult for the disease to expand all
over population.
\vspace{1cm}

\section*{Discussion}
\label{conclu}

In this paper, we studied the propagation of a disease in a population
where all the individuals are continuously debating about getting
vaccinated, considering that a susceptible individual is vaccinated if
he is completely convinced about the benefits of the vaccine. For this
purpose we used two-layer network where in one layer we use the
SIR-model with vaccination for the propagation of a disease, and in
the other layer we used the M-model (with $M=2$), for the opinion
formation process, where compromise and persuasion are the two
processes involved and are controlled by the parameter $r$. We found
that, in all the cases, the number of recovered agents decreases as
$\omega$ increases, and this is due to the fact that as the vaccine
becomes more effective, more people remain vaccinated and the
propagation of the disease slows down. We found an epidemic threshold
$\omega^*$ above which we ensure that an epidemic will not
develop. Furthermore, we found that above a certain value of $\beta^*$
the propagation of the disease is enhanced and it is impossible to
prevent it from becoming an epidemic. Even for $\omega=1$ there will
be a final macroscopic number of recovery individuals in the steady
state. We computed this threshold as a function of $r$, and we found
that a neutral society is the best scenario to prevent an epidemic ($r
\approx 1$). When compromise dominates the process of opinion
formation ($r<<1$), the agents tend to have a moderate opinion, making
difficult that they become extremist in favor of vaccination. The
disease spreads through the non vaccinated agents very easily, even
when the virulence is small. On the other hand, when persuasion
dominates the process of opinion formation ($r>>1$) the agents tend to
have an extremist opinion. All the extremist agents in favor of the
vaccine will be vaccinated, but those agents with a negative opinion,
which are an important fraction of the population, will be easily
infected. In a neutral society it is more likely to convince those
agents with a negative opinion in favor on vaccination, to become
extremist in favor. With compromise and persuasion in the same
proportion it is easier to convince people to get vaccinated, blocking
the propagation of the disease and preventing it to expand all over
the population. We can conclude that the influence of the opinion on
the vaccination determines, in certain cases, whether or not the
disease becomes in an epidemic.

\section*{Acknowledgments}
\label{ackn}

JRI acknowledges financial support of Brazilian agency CNPq and
Argentinian agency CONICET. He also thanks the kind hospitality of the
IFIMAR (Instituto de F\'{i}sica de Mar del Plata) during
2015-2016. LGAZ, CEL and LAB wish to thank to UNMdP, FONCyT and
CONICET (PICT 0429/2013, PICT 1407/2014 and PIP 00443/2014) for
finantial support.

\end{document}